\input{aipcheck}

\documentclass[
    ,final            
  ]
  {aipproc}

\layoutstyle{6x9}

\begin{document}

\title{Hadron spectroscopy at $\bar{\mbox {P}}$ANDA}


\author{Elisa Fioravanti}{
  address={INFN Ferrara - Via Saragat 1, 44122 Ferrara, Italy}
}

\begin{abstract}
FAIR a new International Facility for Antiproton and Ion Reaserach,  is under construction at Darmstadt, in Germany. This will provide scientists in the world with outstanding beams and experimental conditions for studying matter at the level of atoms, nuclei, and other subnuclear constituents. An antiproton beam with intensity up to 2x10$^7$ $\bar{p}/s$ and high momentum resolution will be available at the High Energy Storage Ring (HESR) where the $\bar{\mbox{P}}$ANDA (Antiproton Annihilation At Darmstadt) detector will be installed. In this paper we will illustrate the details of the $\bar{\mbox{P}}$ANDA scientific program related to hadron spectroscopy, after a brief introduction about the FAIR facility and the $\bar{\mbox{P}}$ANDA detector.
\end{abstract}

\maketitle


\section{The FAIR project}

 The Gesellschaft f\"{u}r Schwerionenforschung (GSI) of Darmstadt, Germany, is undergoing a major upgrade of the existing laboratory. This upgrade foresees ion beams of higher intensity and better quality, and, first for GSI, an antiproton beam.\\
The High Energy Storage Ring (HESR) will have two different operation modes: the high intensity mode, where with a beam momentum spread $\delta p/p$=10$^{-4}$ and a luminosity of 2x10$^{32}$ cm$^{-2}$s$^{-1}$ will be available, and the high resolution mode, where the luminosity requirement will be relased to 10$^{31}$ cm$^{-2}$s$^{-1}$ to have a maximum momentum precision of 10$^{-5}$.

\section{The $\bar{\mbox {P}}$ANDA scientific program}

\begin{figure}
  \includegraphics[height=.3\textheight]{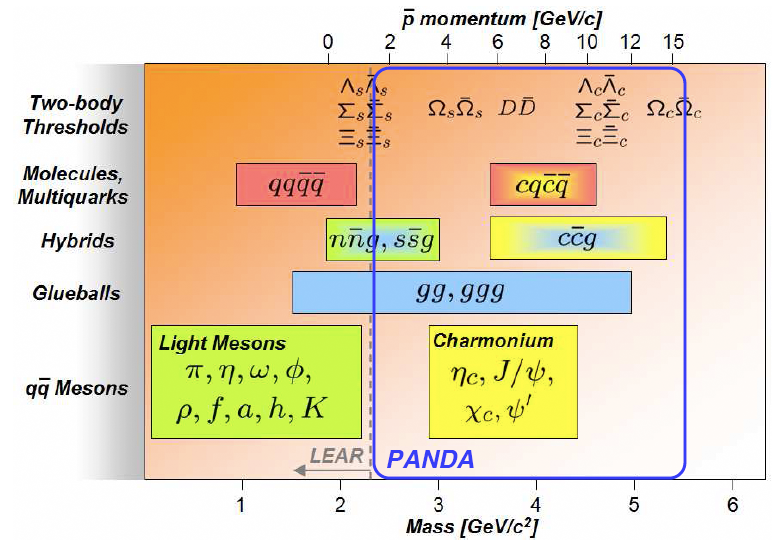}
  \caption{An overview of the various quark and gluon configurations of hadrons and their corresponding mass range. The $\bar{\mbox {P}}$ANDA experiment will exploit masses up to 5.5 GeV/c$^2$ in antiproton-proton collisions, thereby having access to glueballs, charmed hybrids and charmed-rich four-quark states.}
  \label{fig:panda}
\end{figure}

$\bar{\mbox{P}}ANDA$ will use the antiproton beam at the HESR with momentum range between 1.5 GeV/c and 15 GeV/c, corresponding to total center-of-mass energies in the antiproton-proton system between 2.5 GeV and 5.47 GeV. The $\bar{\mbox {P}}$ANDA experiment aims at exploring hadronic matter by means of the gluon rich environment of the $\bar{p}p$ annhilation which allows to access a wide range of final states. The $4\pi$ acceptance of the detector for both charged and neutral particles, together with the envisaged high quality of the antiproton beam, will create an ideal environment to collect high statistics data to address many open problems related with the strong interaction. $\bar{\mbox {P}}$ANDA will perform a complete program of hadron spectroscopy to test many unclear aspects of Quantum Chromo Dynamics (QCD) (Fig. \ref{fig:panda}). The aim is to investigate both the dynamics of the interaction, and the characteristics of new forms of matter such as exotic states in the charm energy range. In the following sections I will focus on charmonium spectroscopy and on the search for non $q\bar{q}$ configurations like glueballs and hybrids.

\section{Spectroscopy of Mesons with Charm}
The charmonium states can be described fairly well in terms of heavy-quark potential models. Precision measurements of the mass and width of the charmonium spectrum give, therefore, access to the confinement potential in QCD. Extensive measurements of the masses and widths of the $\psi$ states have been performed at $e^+e^-$ machines where they can be formed directly via a virtual-photon exchange. Other states, which do not carry the same quantum number as the photon, cannot be populated directly, but only via indirect production mechanisms. This is in contrast to the $p\bar{p}$ reaction, which can form directly excited charmonium states of all non-exotic quantum numbers. As a result, the resolution in the mass and width of charmonium states is determined by the precision of the phase-space cooled beam momentum distribution and not by the detector resolution. \\
The charmonium spectrum is subdivided into two categories by the $D\bar{D}$ threshold. The $c\bar{c}$ states below the $D\bar{D}$ threshold are very narrow and their spectroscopic study requires energy scans across the spectral function in order to determine their central mass and width. Hidden charm states above the $D\bar{D}$ threshold, which do not necessarily have conventional $c\bar{c}$ structure, are much wider and preferentially decay into a $D\bar{D}$ or $D_s\bar{D}_s$ pair. All the $c\bar{c}$ states below the $D\bar{D}$ threshold are known. Yet, a series of question remain to be clarified. For instance, the width of $\eta_c'$ ($2^1S_0$) is known with a very high error \cite{etac2S}. Another open question is the spin-dependence of the $q\bar{q}$ potential. For this, a precise measurement of the mass and decay channels of the singlet-P resonance, $h_c$, is of extreme importance. The available data for the $h_c$ are of poor precision \cite{hc1,hc2}. Due to the narrow width, $\Gamma <$ 1 MeV, of this state, only $p\bar{p}$ formation experiments will be able to measure the width and perform systematic investigations of the decay modes. \\
Above the $D\bar{D}$ threshold the situation is quite different. A number of new states have been recently discovered by BaBar, Belle and CLEO. As so far no deeper understanding of the new states has been achieved, they have been labelled by X, Y and Z. It is clear from the number of these XYZ states and partially also from their properties that a large fraction of them are not conventional charmonium states. One of the most established among the XYZ states is the narrow X(3872) state of less than 2.3 MeV width. It has been discovered by Belle \cite{X3872}. The main hadronic decay modes are $\pi^+\pi^- J/\psi$ and $D^0\bar{D}^0\pi^0$. The narrow width and the decay properties disfavor a $c\bar{c}$ structure. The very close vicinity of the $D^0\bar{D}^{0*}$ threshold favors a molecular interpretation with these constituents. \\
Another state, is the Y(4260), discovered by BaBar as an enhancement in the $\pi\pi J/\psi$ subsystem in the radiative return reaction $e^+e^-\rightarrow\gamma_{ISR}J/\psi\pi\pi$ where "ISR" stands for Initial State Radiation \cite{Y4260}. A number of explanations have appeared in the literature: $\psi(4S)$, $cs\bar{c}\bar{s}$ tetraquark but the hybrid interpretation of Y(4260) is appealing. \\
$\bar{\mbox {P}}$ANDA is ideal to perform such a measurement in direct formation of XYZ states which due to the small beam momentum spread probes the spectral shape with less than 100 KeV resolution.\\
In $\bar{\mbox{P}}$ANDA with a luminosity of $2\cdot 10^{31}$ cm$^{-2}$s$^{-1}$ and assuming 50\% overall efficiency, we expect to have $10^4-10^7$ $c\bar{c}$ states/day.

\subsection{D mesons}

The observation of the new states $D_s(2317)$ and $D_s(2460)$ by Belle, BaBar \cite{D} and CLEO thus came as a surprise, since both states are lower in mass and narrower than expected based on the quark model. In fact the two states are within less than 1 MeV by the same energy gap of 45 MeV below the isospin averaged $DK$ and $D^*K$ thresholds, respectively. An obious interpretation is thus that they could be bound molecules of the respective constituents. In the presentation we concentrate on the $D_s(2317)$. Only an upper limit of its width $\Gamma<$3.8 MeV has so far been measured, whereas theoretical predictions range between a few KeV and about 200 KeV, based on different assumptions on its internal structure. Again $\bar{\mbox {P}}$ANDA can benefit from the good momentum resolution of the antiproton beam to determine the $D_s(2317)$ width performing near-threshold scan. 

\section{Hybrids, glueballs and other exotics}
The self-coupling of gluons in strong QCD has an important consequence, namely that QCD predicts hadronic systems consisting of only gluons, glueballs or bound systems of quark-antiquark pairs with a strong gluon component, hybrids. These systems cannot be categorized as "ordinary" hadrons containing valence $q\bar{q}$ or $qqq$. The additional degrees of freedom carried by gluons allow glueballs and hybrids to have spin-exotic quantum numbers, $J^{PC}$, that are forbidden for normal mesons and other fermion-antifermon systems. States with exotic quantum numbers provide the best opportunity to distinguish between gluonic hadrons and $q\bar{q}$ states. Exotic states with conventional quantum numbers can be identified by measuring an overpopulation of the meson spectrum and by comparing properties like masses, quantum numbers, and decay channels, with predictions from Lattice Quantum Chromodynamics (LQCD) calculations.\\
The first hints for gluonic hadrons came from antiproton annihilation experiment. Two particles, first seen in $\pi N$ scattering with exotics $J^{PC}=1^{-+}$ quantum numbers, $\pi_1(1400)$ \cite{h1} and $\pi_1(1600)$ \cite{h2} are clearly seen in $\bar{p}p$ at rest and are considered as hybrid candidates. In the search for glueballs, a narrow state at 1500 MeV/c$^2$, discovered in antiproton annihilations by the Crystal Barrel collaboration \cite{h3} is considered the best candidate for the glueball ground state $J^{PC}=0^{++}$. However, the mixing with nearby conventional scalar $q\bar{q}$ states makes a unique interpretation difficult.\\
The most promising energy range to discover unambiguously hybrid states and glueballs is in the region of 3-5 GeV/c$^2$, in which narrow states are expected to be superimposed on a structureless continuum. In this region, LQCD predicts an exotic $1^{-+}$ $\bar{c}c$-hybrid state with a mass of 4.2-4.5 GeV/c$^2$ and a glueball state around 4.5 GeV/c$^2$ with an exotic quantum number of $J^{PC}=0^{+-}$ \cite{h4}. The $\bar{p}p$ production cross section of these exotic states are similar to conventional states and of the order of 100 pb. All other states with ordinary quantum numbers are expected to have cross sections of about 1 $\mu$b.


%

\section{Summary}

The  $\bar{\mbox {P}}$ANDA  experiment at FAIR will address a wide range of topics in the field of QCD, of which only a small part could be presented in this paper. The physics program will be conducted by using beams of antiprotons together with a multi-purpose detection system, which enables experiments with high luminosities and precision resolution. This combination provides unique possibilties to study hadron matter via precision spectroscopy of the charmonium system and the discovery of new hadronic matter, such as charmed hybrids or glueballs, as well as by measuring the properties of hadronic particles in dense environments. New insights in the structure of the proton will be obtained by exploiting electromagnetic probes. 


\end{document}